\documentclass[twocolumn,aps, prl]{revtex4-2}
\usepackage{graphicx}
\usepackage{amsmath}
\usepackage[colorlinks]{hyperref}
\usepackage{color}

\begin{document}

\title{Detecting infected asymptomatic cases in a stochastic model for spread of  Covid-19. The case of Argentina.}

\author{N.~L.~Barreiro}
\affiliation{\small Instituto de Investigaciones Científicas y Técnicas para la Defensa (CITEDEF), Buenos Aires, Argentina.}

\author{T.~Govezensky}
\affiliation{\small Instituto de Invesitgaciones Biomédicas, Universidad Nacional Autónoma de México, Mexico 04510, México.}

\author{P.~G.~Bolcatto}
\affiliation{\small Instituto de Matemática Aplicada del Litoral (IMAL, CONICET/UNL), FHUC. Santa Fe,3000, Argentina}
\affiliation{\small Instituto de Investigaciones Científicas y Técnicas para la Defensa (CITEDEF), Buenos Aires, Argentina.}

\author{R.~A.~Barrio}
\affiliation{\small Instituto de Física. Apartado Postal 20-365, Universidad Nacional Autónoma de México, Mexico 04510, México.}

\date{\today}

\begin{abstract}
We have studied the dynamic evolution of the Covid-19 pandemic in Argentina. The marked heterogeneity in population density and the very extensive geography of the country becomes a challenge itself. Standard compartment models fail when they are implemented in the Argentina case. We extended a previous successful model to describe the geographical spread of the AH1N1 influenza epidemic of 2009 in two essential ways: we added a stochastic local mobility mechanism, and we introduced a new compartment in order to take into account the isolation of infected asymptomatic detected people. Two fundamental parameters drive the dynamics: the time elapsed between contagious and isolation of infected individuals ($\alpha$) and the ratio of people isolated over the total infected ones ($p$). The evolution is more sensitive to the $p-$parameter. The model not only reproduces the real data but also predicts the second wave before the former vanishes. This effect is intrinsic of extensive countries with heterogeneous population density and interconnection.The model presented here becomes a good predictor of the effects of public policies as, for instance, the unavoidable vaccination campaigns starting at present in the world.

\end{abstract}


\maketitle

\section{Introduction}
\label{sec:intro}


The pandemic of 2020 has changed life in many respects. In the scientific community there have been an increased interest in all the issues related to this phenomenon. More than 23500 articles published containing the word Covid19 this year account for this fact. There are studies on the medical and biological aspects of the virus, the mechanisms of contagion, the strategies to avoid the spread of the disease, the recommendations of health authorities to prevent infection as to keep a respectable distance for others, the use of masks, washing hands profusely and frequently, sanitizing shoes and utensils, and compulsory test and tracking techniques \cite{WHO:2020}. Total confinement has been applied since the 14th century {\cite{Gensini:2004}, it is obviously} effective because no personal contact means no infection spread, but this option is not realizable in the modern societies for a long period of time. A recent action that has proved useful is to detect  not only infected, but also asymptomatic people by random testing and isolate the ones that come out positive \cite{Islandia}. 


When a new virus emerges, and there is no effective treatment or vaccine, these non-pharmacological interventions (NPIs) constitute the main response option for mitigating the effects of the pandemic. Search - based on observational data and mathematical models - has been done to identify effective NPIs, one of them is quarantine, and its effectiveness increases when applied early in the pandemic and in combination with other NPIs \cite{Nussbaumer:2020, Odusanya:2020}. Active search for infected people who are asymptomatic or present mild symptoms and subsequent isolation was not included in the above mentioned research, probably because few countries had implemented it. However, for SARS-Cov-2, a great proportion of infected people are asymptomatic or present mild symptoms \cite{Pollan:2020,Stringhini:2020}, they are not quarantined, still they remain infectious until they are recovered \cite{Salje:2020}. More research is needed to further assess the effect of this kind of procedure
Then, it becomes necessary to model the spread of the pandemic in this situation and compare it with predictions without infected asymptomatic people detection.


Compartmental models have been extended to include quarantined individuals \cite{Castillo:2003, Hethcote:2002, Erdem:2017, ElFatini:2020}. Although time dependent parameters have been used to simulate people’s mobility \cite{Tagliazucchi:2020},
most of the implementations do not include the actual geographical spread, assuming in general homogeneous populations and equally probable interactions among individuals. However, when modeling pandemics, demographic heterogeneity, and people mobility could be a key element to be considered. In particular, human mobility is strongly affected by governments' policies and it is almost imperative to include this feature in order to render simulations closer to the real data.

We propose here a model in which the noisy interactions of human society and the inhomogeneities of geographical spread are taken into account. We use an extension of a model proposed in Ref. \cite{Barrio:2009} to predict the influence of these measures of detection and subsequent isolation. The original model has been extremely successful in predicting the behavior of Covid-19 in various countries, as different in all respects as Mexico, Finland, and Iceland \cite{Barrio:2020}. The dynamic works in two scales: On the one side a micro or local one in which the disease spread follow a (almost) standard compartmental evolution. It contains the biology-related parameters of the disease. On the other side, a macro or long range dynamics, which describes the geographical interconnection in a given region or country. However, the implementation as in \cite{Barrio:2020} was surprisingly inaccurate for the Argentina case. To understand this singular behavior we have to give more versatility to the model by incorporating the influence of detecting and isolating infected asymptomatic persons as well as to account for local stochastic mobility.

In the next Section we describe the model in detail, then we explain how it is applied to the Argentina case, and then present some results from numerical calculations. Finally we conclude with some important remarks.

\section{Theoretical model}
\label{sec:modelo}

The approach is based on a collection of SEIR models acting in cells distributed along (and filling) the whole geography of the country. The network is weighted by the population density of each cell. Connections between cells are realized by the national ground and/or air roads. This approach has the advantage that the parameters proper to the disease (only in the SEIR part of the model) are separated from the ones related to spreading infections between people, which ultimately translate into mobility quantities between cells.

\subsection{ Local dynamics: SEIQR stochastic model (Mycrodinamics)}



The SEIR model of Ref. \cite{Barrio:2009} was converted into a SEIQR stochastic model in order to analyse the influence of detection and isolation of infected people. This was done by adding a quarantine compartment, $Q$.
Fig. 1 shows a diagram of the five-compartmental model: susceptible individual ($X$), exposed yet not infectious ($E$), infectious ($I$), isolated in quarantine ($Q$), and deaths or recovered while acquired immunity lasts ($Z$). After the immunity period, people could be again susceptible to the disease according to the survival parameter $S$. Incubation period ($\epsilon$), infectiousness period ($\sigma$ ) and immunity period ($\omega$) are parameters depending on the specific disease studied and host’s immune response. Two new parameters are added to the model: $\alpha$ - the time lapse from infection to detection and isolation of infected individuals, and $p$ - the proportion of infected people detected and put in quarantine, either symptomatic or asymptomatic. All time parameters are assumed to be constant and dimensionless by expressing them in units of a time scale of one day.

\begin{figure}[h!]
\centering
\includegraphics[width=1.0\columnwidth]{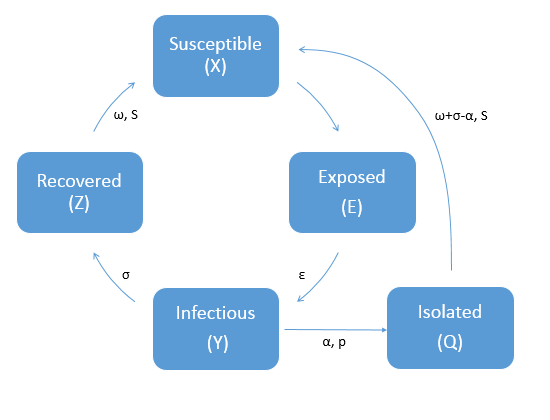}
\caption{Compartment scheme of a SEIQR model. $\epsilon$, $\alpha$, $\sigma$ and $\omega$ are the latency, isolation, infectiousness and immunity periods, $p$ is the portion of infected people which is isolated and  $S$ is the survival parameter. }
\label{fig:esquema}
\end{figure}


Demography is added using a constant mortality rate $\mu$=1/$L$, where $L$ is life expectancy. The total population $N=X+E+Y+Q+Z$ depends only on the survival parameter $S$; at time zero $X=N$.
In order to keep $N$ fixed, the birth rate is considered a constant ($N\mu$) and all the newborns are included in the susceptible compartment. Since we assume that the disease is spread due to daily contacts, we consider that the dynamical evolution of the model should be given in discrete steps of one day, which is the time unit of all the delay parameters.

With all these assumptions, the discrete SEIQR model could be written as discrete mathematical map with five variables,
\begin{widetext}
\begin{eqnarray}
\label{ec:modelo}
 X_{t+1} &=& (1-\mu)~(X_{t}- G_{t}+ S (1-\mu)^{\epsilon +\sigma + \omega} G_{t-1-\epsilon -\sigma - \omega}) +\mu N \\
 E_{t+1} &=& (1-\mu) (E_{t}+ G_{t} - (1-\mu)^{\epsilon} G(t-1-\epsilon ))\\
 Y_{t+1} &=& (1-\mu)~(Y_{t}+ (1-\mu)^{\epsilon} G_{t-1-\epsilon}- (1-p)~(1-\mu)^{\epsilon+\sigma} G_{t-1-\epsilon -\sigma }-p~(1-\mu)^{\epsilon+\alpha} G_{t-1-\epsilon -\alpha })\\
 Q_{t+1} &=& (1-\mu)~(Q_{t}+p~(1-\mu)^{\epsilon+\alpha} G_{t-1-\epsilon -\alpha }-p~(1-\mu)^{\epsilon+\sigma + \omega} G_{t-1-\epsilon -\sigma - \omega} )\\
 Z_{t+1} &=& (1-\mu)~(Z_{t}+(1-p)~(1-\mu)^{\epsilon +\sigma } G_{t-1-\epsilon -\sigma}- (1-p)~(1-\mu)^{\epsilon +\sigma + \omega} G_{t-1-\epsilon -\sigma - \omega})\\ 
\end{eqnarray}
\end{widetext}
where $G_t$ is the incidence function evaluated at time $t$. 
Assuming a homogeneously mixed population within a cell, the probability of getting infected is calculated using the Poisson probability distribution; then the incidence rate is given by:
   $G_t = X_t (1- e^{- \beta Y_t}),$
where $\beta$ is a transmission parameter that characterizes the intrinsic behavior of the disease and it is a dimensionless constant for a specific infection. We assume $G_t=0$ for $t < 1$.


A model of this type is acting in each cell of the geographical region of interest. The described system of equations is deterministic. However, since people move not only to fixed places, following daily routine activities, but also may go to nearby  unpredictable places, random mobility within a cell should also be considered  \cite{Bittihn:2020}, which we denote as {\it local mobility}. This is done by adding a threshold parameter $v_L$, between 0 and 1, that accounts for people's short distance mobility. Each day $t$, in every geographical cell with coordinates $(i; j)$, we compare a random number ($r$, from uniform probability distribution between 0 and 1) with this threshold parameter. If $v_L<r$ the systems keeps its normal evolution but, if $v_L>r$ we regard that the mobility was low so that the disease didn't evolve; under this circumstance, in this area, there are not new infected people during this day.

\subsection{Geographical disease spreading (Macrodynamics)}


 A realistic model of epidemics most include its geographical spread in big regions or counties with heterogeneous population densities. For this purpose, the map of the country under study is divided in a two dimensional grid of squares of size of a few km$^2$. For each cell of coordinates $(i; j)$, the actual population density $\rho(i; j)$ is known. Within each cell population is normalized to $N=1$, local dynamics is simulated by a SEIQR stochastic model using the incidence function weighed by its population density:

\begin{align}
   G_t(i,j) = X_t(i,j) \rho(i,j) (1- e^{- \beta Y_t(i,j)})
   \label{ec:incidence1},
\end{align}    
      
To consider the spread among first neighbor regions we use a Metropolis Monte-Carlo algorithm. For each square in the grid, if $Y_t(i,j)$ $\leq$ $\eta$ and  $\nu_n <r $, there is propagation of the disease to a neighbor cell. The value $\eta$ is related to the infectiousness of the disease and $\nu_n$ varies between 0 and 1 and accounts for the mobility between neighbors. $r$ is a random number given by a uniform distribution between 0 and 1. To start the disease in a new cell of coordinates $(i,j+1)$, which is one first neighbor of the cell $(i,j)$, the initial conditions are given by $Y_t(i,j+1)=\eta$ and $X_t(i,j+1)=1-\eta$.
The disease can also be spread randomly to distant regions because of people traveling between connected cities. Another Metropolis Monte-Carlo algorithm is used for long distance new infections, either by road or by air.
It is more likely that bigger cities are infected first because they are more populated and connected. Because of this, the long distance mobility parameter $\nu_a$ is weighed by the normalized densities of both, origin and destiny cells. In this case propagation occurs between a cell already infected ($Y_t(i,j)$ $\leq$ $\eta$) and a cell connected to the first one by air, trains or national routes. If  $\nu_a \rho(i,j) \rho(m,n) <r $, (with $r$ a random number from a uniform probability distribution between 0 and 1) the cell at the new coordinates (m,n) starts the disease with initial conditions $Y_t(m,n)=\eta$ and $X_t(m,n)=1-\eta$.

Finally, since people move in an essentially random way it is possible to find people travelling to distant cities or even isolated towns with lower population densities. This is accounted for  by the noise parameter $KT$ representing
the "kinetic energy" of the system. In this case a new Monte-Carlo algorithm is applied. For each cell with coordinates $(i,j)$ with $\rho(i,j)> T$, (with T a normalized population density threshold), if $e^{-1/KT}<r $, then the disease will start at the cell $(i,j)$ with initial conditions $Y_t(m,n)=\eta$ and $X_t(m,n)=1-\eta$.


In this model $\beta$ does not depend on $\rho$ and/or on mobility of people as in traditional SEIR models.  $\rho$ is considered constant throughout the pandemic, and mobility parameters can be used to reflect measures applied by different governments trying to control the pandemic. Another advantage of this model is that since detected individuals, and therefore reported, are clearly separated from non-reported asymptomatic ones, the model traces both groups, an important point in the case of SARS-CoV-2 where there are so many asymptomatic people.

\section{Application of the model to COVID-19 in Argentina}
\label{sec:Aplicacion}


In order to apply these ideas to the case of Argentina we have divided the country's territory in a grid of around 67000 squares of 7 km $\times$ 7 km. The total population inside of each parcel was assigned from the data provided by the National Geographic Institute of Argentina (IGN). 
The interconnection between cities by commercial flights was canceled by public policies since the early days of the pandemic. Consequently, only land connections are possible. Therefore, we allow traveling (both, short and large distances) across the network of roads and routes also provided by the IGN.

\begin{figure}[ht]
\centering
\includegraphics[width=0.5\columnwidth]{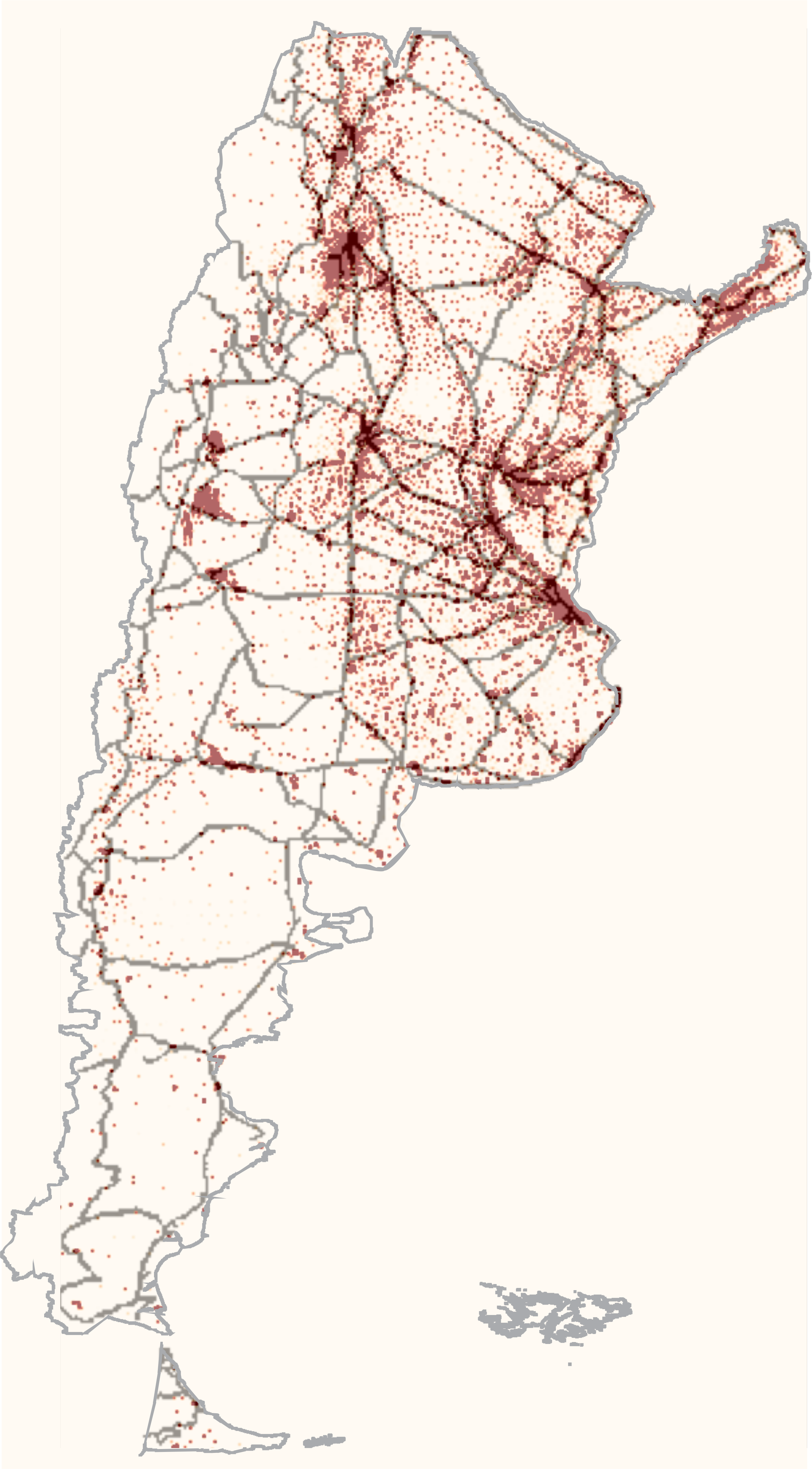}
\caption{Density map and distribution of routes used in the model. The information to create the maps was provided by the IGN. Each pixel corresponds to a 7 km x 7 km parcel}
\label{fig:ejemplo00}
\end{figure}

\subsection{Fitting parameters to Argentina}

It is important to remark that in this model the disease parameters are well separated from those that account for the social distancing and mobility. Since there is not much information about latency, infectiousness and immune periods of COVID-19, and this values could vary significantly from person to person, one can fit this values to the actual data within upper and lower thresholds found in the literature \cite{CDC:AUG2020}.

The first confirmed case of COVID-19 in Argentina was a person coming from abroad on March 3th, 2020. And, in general, the disease evolution during the first 20 days was because of infected people coming from Europe and USA and not because of community spread. On March 20th a strict lock-down started all around the country and mobility was drastically reduced. Because of this, disease parameters cannot be fitted straight from Argentina data. In this sense we decided to use the same parameters as in México, Iceland and Finland models \cite{Barrio:2020}:  $\epsilon=1$, $\sigma=14$ and $\beta=0.91$. $\omega$ was conservatively chosen as 140 days because the center for disease control and prevention (CDC) states that there are no reports of people being reinfected within 5 months of first infection\cite{CDC:AUG2020} . $\sigma$ is set to the quarantine standard time used in several countries (Argentina among them). 
 At present, there are scarcely any available studies in Argentina to sure how much infected people is detected and put in quarantine (quantified in our model by the $p-$ parameter). One particular study in urban slum dwellers of Buenos Aires City suggests that only 10$\%$ of the actually infected peoples was PCR tested and nearly 90$\%$ were asymptomatic cases \cite{Figar:2020}. Patients are tested only when presents two or more concurrent symptoms and, in the case of a positive result, all of the close contacts are isolated independently of new tests. Since some studies suggest that most COVID-19 patients have mild symptoms or are asymptomatic \cite{Eskild:2020,Oran:2020,Alvin:2020}, we estimate that tested and isolated cases in Argentina are between 10$\%$ and 20$\%$ of the actual number of infected people, i.e. $0.1 \leq p \leq 0.2$. These values are similar in order of magnitude to those found in Spain and France \cite{Pollan:2020,Salje:2020}. To estimate $\alpha$ we use the information provided by the Health Ministry of Argentina for each patient.
 From this data, we found that 80$\%$ of the people start the isolation between 3 and 7 days after they get infectious although mostly at 5 days. Since this is a mean value model, we decided to use 5 as a good estimation.  Conservatively, we choose the values $p=0.1$ and $\alpha=5$ to study the evolution of the pandemic in Argentina.

For simplicity, mobility parameters are regarded equal $\nu_a$=$\nu_l$=$\nu_n$=$\nu$. Consequently they change on time in the same way. Since mobility has drastically reduced during the first stages of the pandemic, the noise was considered very low taking a value $KT=0.1$. We take four intervention times to account for the changes in the stringency of the government measures. This intervention dates were fitted with a delay of around 7-8 days with the real implementation dates because we noticed that this is, in average, the time taken by a measure to impact in the growth rate of the pandemic. The government measures are described in Table \ref{tab:table1} 



\begin{table*}
\caption{\label{tab:table1} Mobility parameters fitted for $p=0.1$ and $\alpha = 5$}
\begin{ruledtabular}
\begin{tabular}{ccc}
 Period (days)  & $\nu$ & Government Measures \\
\hline
 1 to 22  & 0.33 
 & Schools and mass attendance events were closed and people were asked to stay at home. \\
 23 to 78 &  0.135  & Strict lock-down in all the territory \\
 79 to 101 &	0.185 & Banks and other businesses were allowed \\
 102 to 165 &	 0.225 & After a massive strike the pandemic evolution raised consistently. \\ &&The lock down started to be lifted.\\ 
 166 - present  &	0.358 & Most restrictions were gradually lifted. The pandemic spread over all the country.\\
\end{tabular}
\end{ruledtabular}
\end{table*}

The $\eta$ parameter was fixed to $10^{-5}$, which corresponds to start the disease in a city with at least 10 infected people. 

Finally, considering a 2.7$\%$ case fatality rate, we set the survival parameter to $S=0.9973$ (fatality rate depends on the amount of tested cases).

\begin{figure}[ht]
\centering
\includegraphics[width=1\columnwidth]{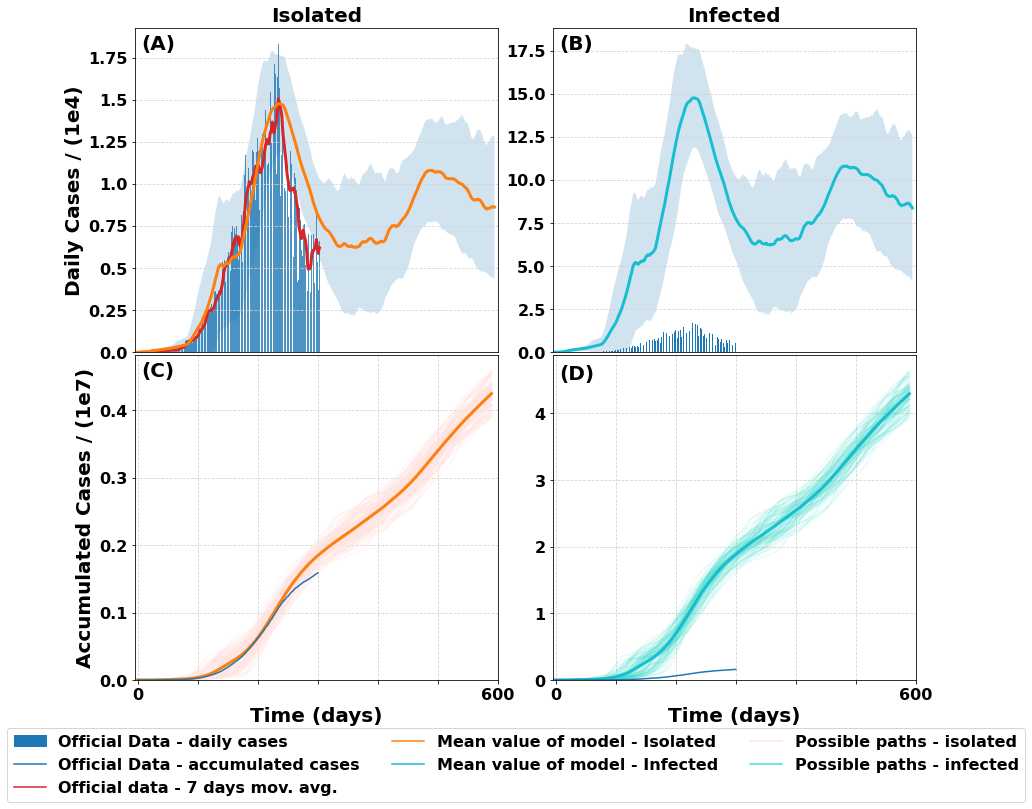}
\caption{Geographical spread stochastic SEIQR Model fitted to Argentine data for $p=0.1$ and $\alpha=5$. (A) Daily isolated cases compared to official data. The blue shaded area corresponds to one standard deviation of the mean. (B) Daily infected cases compared with official Data. (C) Mean Value of 100 model runs- isolated.  (D) Mean Value of 100 model runs- infected.  }
\label{fig:ejemplo1}
\end{figure}

\subsection{Comparison with reported data}
In Fig. \ref{fig:ejemplo1} we show the result of this model obtained from adding the newly isolated people from all the cells in the grid scaled by each region's own population. This plot is obtained by averaging 100 model runs. We should keep in mind that most governments are not able to detect all the actual COVID-19 cases but a fraction $p$ and that only the confirmed and isolated people should be compared with the data provided by official sources.   

In order to understand Fig. \ref{fig:ejemplo1}, three points should be kept in mind. First one is that the mobility parameter from day 166 onward was fixed to 0.358. This means that this prediction will be adequate as long as the people keeps respecting social distancing measures. It should be regarded that if all the mobility and group meeting restrictions are lifted the evolution will be different. The second topic to consider is the appearance of a second pandemic wave. This wave is strongly related with the immunity period which was fixed at 140 days. Since there is not enough data available at this time it is possible that this second wave appears a few weeks earlier or later. Obviously, this dynamic will be dramatically different if a campaign of massive vaccination occurs. 

The third issue to notice is that, as it is expected, the infected are 10 times higher than the isolated as it can be observed by comparing figures (A) and (B) or (C) and (D). At the current transmission rate, this implies that by the end of 2021 the pandemic will have infected a number of people equivalent to the country's population. If we compute fatalities as the 0.027\% of the infected, we can predict that there will be 116.000 deaths by the beginning of 2022.

\begin{figure}[ht!]
\centering
\includegraphics[width=1\columnwidth]{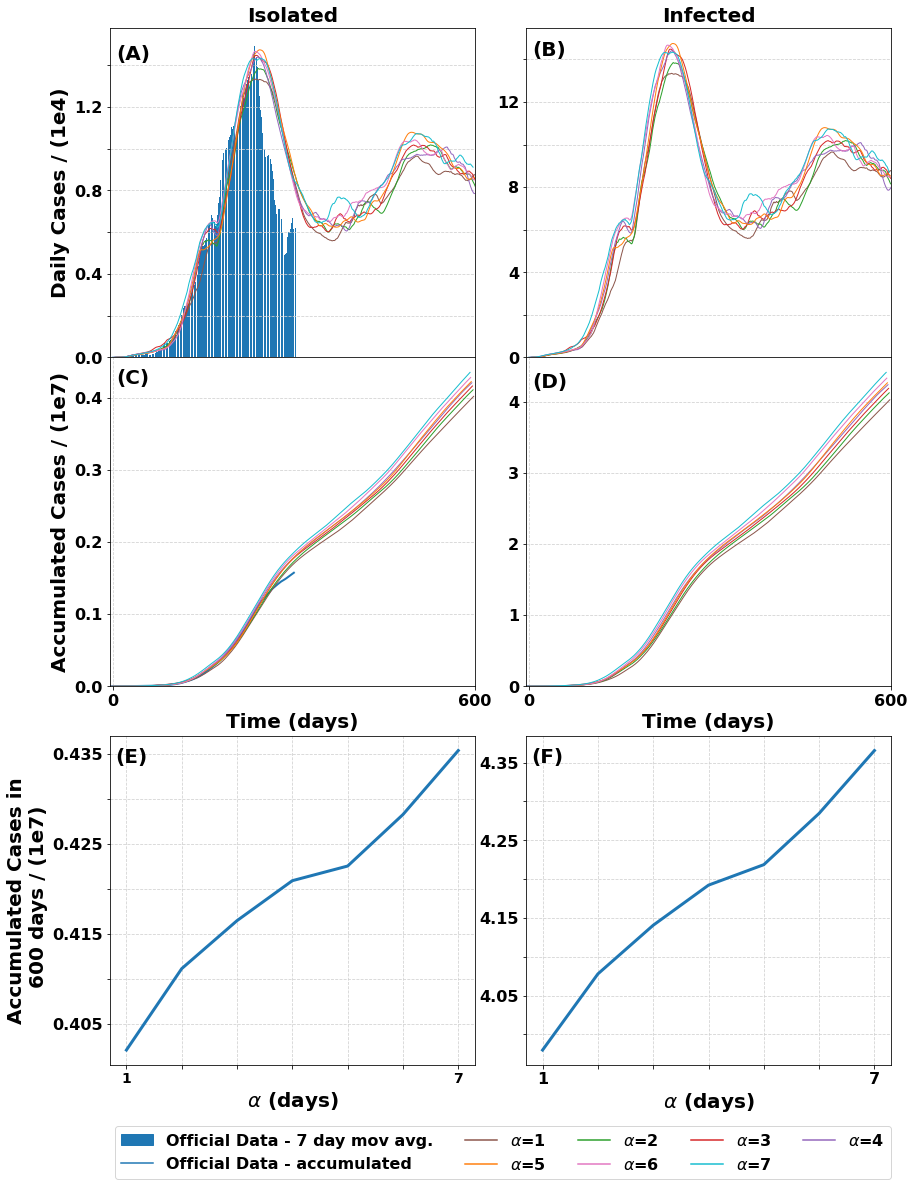}
\caption{ Model prediction according to different values of $\alpha$. For each curve we did 100 runs of the model. The value $p$ is fixed to $0.1$ and the mobility is the one fitted for $\alpha=5$. (A)Daily discovered cases. (B)Daily actually infected cases. (C) Accumulated discovered cases. (D) Accumulated infected cases. (E) Accumulated discovered cases on 600 days. (F) Accumulated infected cases on 600 days.}
\label{fig:ejemplo2}
\end{figure}

In order to study how the $p$ and $\alpha$ parameters affect the evolution of the pandemic, we analyzed different combinations of them and fixing the values of $\nu$ as in the Table \ref{tab:table1}. 
Firstly, we study the model by varying $\alpha$. Figure \ref{fig:ejemplo2} shows the curves obtained from 100 runs of the model for $p=0.1$ and different values of $\alpha$. As it is clear from the figure, the early discovery of a case reduces the height and the width of the peak in (A) and (B) but not in a significant way. This can be seen clearly in the curves (E) and (F) were the accumulated isolated and infected cases in 600 days are shown. The expected difference between early and late case discovery for $ p = 0.1 $ is around 8\%, which is within the spread of the model. The value taken by $\alpha$ could become more significant for bigger $p$ but, as we have seen before, the fraction of discovered and isolated people is small in most countries. 

\begin{figure}
\centering
\includegraphics[width=1\columnwidth]{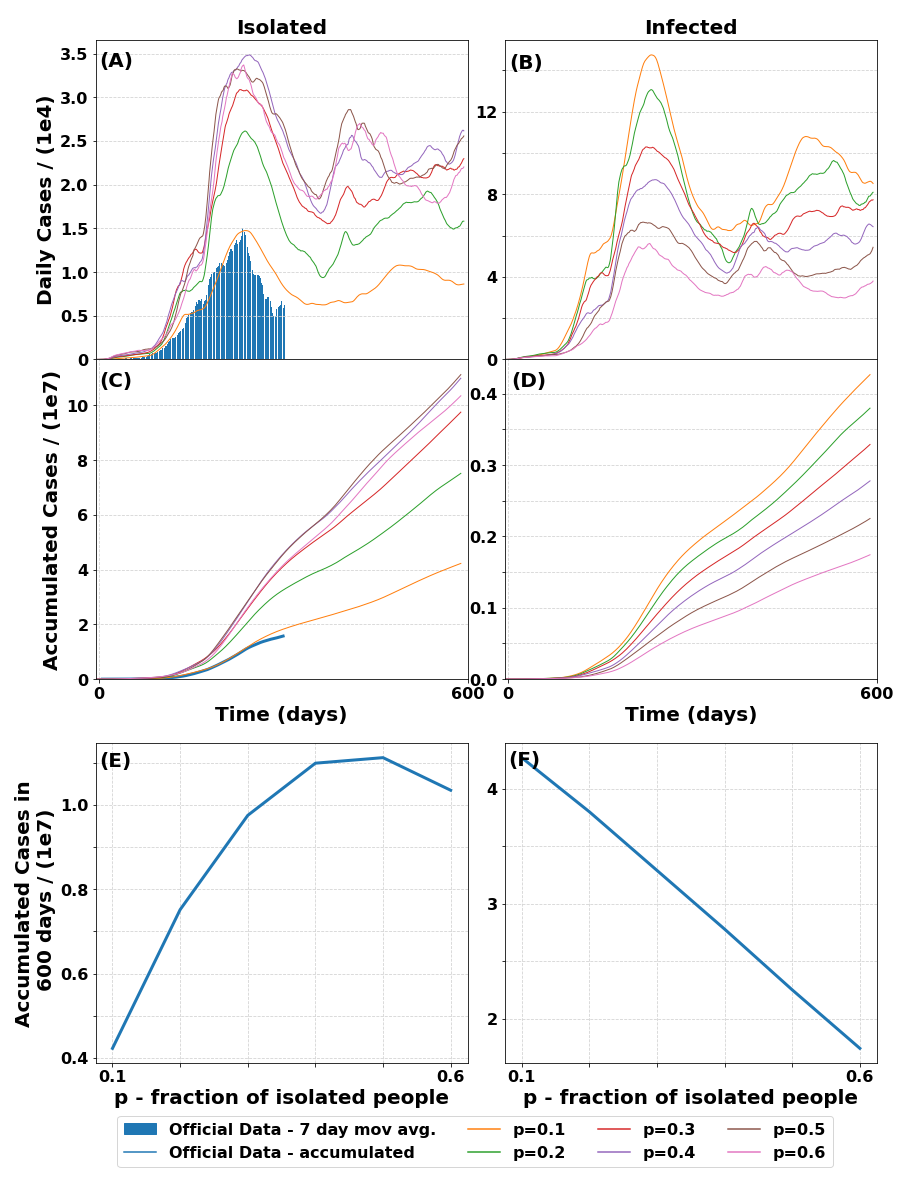}
\caption{Model prediction according to different values of $p$. For each curve we did 100 runs of the model. The value $\alpha$ is fixed in 5 days and the mobility is the one fitted for $p=0.1$ (A)Daily discovered cases. (B)Daily actually infected cases. (C) Accumulated discovered cases. (D) Accumulated infected cases. (E) Accumulated discovered cases on 600 days. (F) Accumulated infected cases on 600 days. }
\label{fig:ejemplo3}
\end{figure}

Next we studied the variation of the fraction of isolated people, $p$. The results can be seen in Figure \ref{fig:ejemplo3}. It is clear from the figures (D) and (F) that the accumulated infected are smaller for smaller $p$.This result is reasonable since the higher the fraction of discovered infected people, the smaller is the population that continues to infect others reducing the spread of the pandemic. This shows that the implementation of an efficient COVID-19 tracking and testing program could be of great significance to control its evolution.

Figures \ref{fig:ejemplo3} (C) and \ref{fig:ejemplo3} (E) show the number of accumulated infected cases. In this case, as the number of  discovered cases raises, the infected population decreases leaving a smaller pool of people with the virus to be found. Therefore, once that more than 40 \% of those infected are discovered, a decrease in the number of isolated cases is observed as a consequence of the reduction in the total diseased population. It is interesting to notice that, for the same parameters and mobility, we found that the total accumulated infected cases in 600 days is 57\% smaller for $p = 0.6$ than for $p = 0.1$. This would imply a reduction in the expected fatalities to less than 50.000.

\begin{figure}
\includegraphics[width=1\columnwidth]{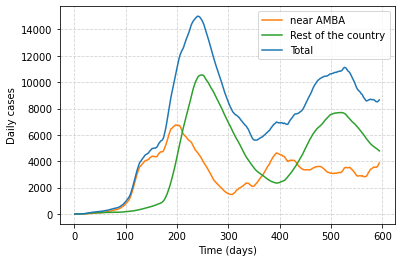}
\caption{Geographical evolution of the pandemic. The model shows a 50 days delay between the peak near-AMBA region (Buenos Ares city and the nearest urban conglomerates, not all the metropolitan area) and the rest of the country. This fact prevent the complete suppression of the disease.}
\label{fig:ejemplo6}
\end{figure}

We want to highlight the use of this new model to accurately describe the geographical spread of the pandemic in the Argentinean territory. From all the figures it can be clearly seen that, without any government measures as mobility reductions, effective infectious tracking or vaccines, the disease is not expected to disappear by itself. The model also predicts the appearance of a second wave before the end of the first one. This is a direct consequence of the geographical extension of the territory and the stochasticity of the model. When the pandemic starts in a certain region of the country also diminishes in another area at the same time (see figure \ref{fig:ejemplo6}). In this way, the disease oscillates between different regions delayed in time. Consequently, it never completely stops or vanishes. The $\sim$50-day elapsed between maxima are very close to that actually seen in the official data. The illness started mainly in the metropolitan area of Buenos Aires (AMBA), which concentrates almost 33\% of the population of Argentina. After several months of evolution, the pandemic moved to other important urban areas of the country as Córdoba, Santa Fe, Río Negro, Mendoza, Chaco, etc., and started diminishing in the AMBA region. The great mobility between these areas will eventually create a new peak in the AMBA region once the immunity period is finished for most of its population.

This kind of geographic behaviour can also be seen in countries like USA, Brazil and Mexico because they have big territories with several important urban areas. As an example, USA had its first cases peaks in the North east states during April and then the pandemic moved to south western states where the cases peaks where on July. Because of this it never experienced an overall decrease in the amount of cases as in most European countries and, as a consequence, they are facing the second wave in states like New York without reaching the end of the first wave in other states. 

\section{Conclusions}
In this work we have proposed a novel model to study the influence of geographical and sociological conditions on the spread of the virus SARS-Cov2, causing the Covid-19 pandemic in Argentina. 

We introduce two fundamental parameters: $\alpha$, the time elapsed between contagious and isolation of infected individuals, and $p$, the ratio of people isolated over the total infected ones. The results show clearly that the detection and consequent isolation of infected people are crucial to the dynamics of the virus propagation ($p-$sensitivity) while the time lasted between infection and isolation is not a relevant issue ($\alpha-$insensitivity).

We also introduced local mobility into the model to account for random social behavior within each cell. This allowed a better prediction of the pandemic evolution and enabled the appearance of new features that were actually observed in real data. For instance, the model predicts new waves whose period depends mainly on the immunity time parameter ($\omega$). This feature was already observed in European countries where the second wave appeared and reinfection was observed. Moreover, the model shows an interesting behavior in countries with a wide geographic span and high mobility. The new daily cases are expected to come from different regions at different moments as the pandemic evolves. This creates unsynchronized oscillations of the daily case curves for different areas and prevents the disease from being completely eradicated. In order to stop this kind of behavior, a better control of the mobility between distant regions should be adopted. Testing and quarantine policies already embraced in some countries could prevent the appearance of new infectious foci.

As final remarks, we want to emphasize the importance of tracking and isolating infected people. In this sense, countries as Iceland and South Korea have shown the effectiveness of these methods to reduce the pandemic spread \cite{Barrio:2020, Park:2020}. On the other hand, cultural habits and social behavior have shown to be important factors as well. The increase in the number of cases because of the mobility growth was clearly demonstrated in the fitted $\nu$ parameters. With this in mind, social distancing measures and case tracking are two key factors to contain the pandemic evolution. Furthermore, as this model shows robustness as a global predictor in a very extensive and heterogeneously connected country like Argentina, we are confident that it gives strong support to analyze vaccination strategies for the future mitigation of the disease.

\section*{Acknowledgments}

RAB and NLB acknowledge support from The National Autonomous University of Mexico (UNAM) and Alianza UCMX of the University of California (UC), through the project included in the Special Call for Binational Collaborative Projects addressing COVID-19. RAB was financially supported by Conacyt through project 283279.

\bibliography{seiirs}

\begin{thebibliography}{99}

\bibitem{WHO:2020}
World~Health Organization.
\newblock {\em Coronavirus disease (COVID-19) advice for the public [ONLINE]},
  2020.
\newblock
  \url{https://www.who.int/emergencies/diseases/novel-coronavirus-2019/advice-for-public}.

\bibitem{Gensini:2004}
Gian~Franco Gensini, Magdi~H. Yacoubc, and Andrea~A. Conti.
\newblock {The concept of quarantine in history: from plague to SARS}.
\newblock {\em Journal of Infection}, 49:257--261, 2004.

\bibitem{Islandia}
{\em Official information about COVID-19 in Iceland}, 2020.
\newblock \url{https://www.covid.is/sub-categories/iceland-s-response}.

\bibitem{Nussbaumer:2020}
Nussbaumer-Streit B., Mayr V., Dobrescu AI., Chapman A., Persad E., Klerings
  I., Wagner G., Siebert U., Christof C., Zachariah C., and Gartlehner G.
\newblock {Quarantine alone or in combination with other public health measures
  to control COVID-19: a rapid review.}
\newblock {\em Cochrane Database of Systematic Reviews}, page Art. No.:
  CD013574, 2020.

\bibitem{Odusanya:2020}
Odusanya O.O., Odugbemi B.A., Odugbemi T.O., and Ajisegiri W.S.
\newblock {COVID-19: A review of the effectiveness of nonpharmacological
  interventions.}
\newblock {\em Niger Postgrad Med J}, 27:261--7, 2020.

\bibitem{Pollan:2020}
Marina Pollán, Beatriz Pérez-Gómez, Roberto Pastor-Barriuso, Jesús Oteo,
  Miguel~A Hernán, Mayte Pérez-Olmeda, Jose~L Sanmartín, Aurora
  Fernández-García, Israel Cruz, Nerea~Fernández de~Larrea, Marta Molina,
  Francisco Rodríguez-Cabrera, Mariano Martín, Paloma Merino-Amador,
  Jose~León Paniagua, Juan~F Muñoz-Montalvo, Faustino Blanco, Raquel Yotti,
  and on~behalf of~the ENE-COVID Study~Group.
\newblock {Prevalence of SARS-CoV-2 in Spain (ENE-COVID): a nationwide,
  population-based seroepidemiological study}.
\newblock {\em The Lancet}, 396:535--544, 2020.

\bibitem{Stringhini:2020}
Silvia Stringhini, Ania Wisniak, Giovanni Piumatti, Andrew~S Azman, Stephen~A
  Lauer, Hélène Baysson, David~De Ridder, Dusan Petrovic, Stephanie
  Schrempft, Kailing Marcus, Sabine Yerly, Isabelle~Arm Vernez, Olivia Keiser,
  Samia Hurst, Klara~M Posfay-Barbe, Didier Trono, Didier Pittet, Laurent
  Gétaz, François Chappuis, Isabella Eckerle, Nicolas Vuilleumier, Benjamin
  Meyer, Antoine Flahault, Laurent Kaiser, and Idris Guessous.
\newblock {Seroprevalence of anti-SARS-CoV-2 IgG antibodies in Geneva,
  Switzerland (SEROCoV-POP): a population-based study}.
\newblock {\em The Lancet}, 396:313--319, 2020.

\bibitem{Salje:2020}
Henrik Salje, Cécile~Tran Kiem, Noémie Lefrancq, Noémie Courtejoie, Paolo
  Bosetti, Juliette Paireau, Alessio Andronico, Nathanaël Hozé, Jehanne
  Richet, Claire-Lise Dubost, Yann~Le Strat, Justin Lessler, Daniel Levy-Bruhl,
  Arnaud Fontanet, Lulla Opatowski, Pierre-Yves Boelle, and Simon Cauchemez.
\newblock {Estimating the burden of SARS-CoV-2 in France}.
\newblock {\em Science}, 369:208--2011, 2020.

\bibitem{Castillo:2003}
Castillo-Chavez C., Castillo-Garsow C., and Yakubu A.
\newblock Mathematical models of isolation and quarantine.
\newblock {\em J Am Med Assoc}, 290:2876--2877, 2003.

\bibitem{Hethcote:2002}
Herbert Hethcote, Ma~Zhien, and Liao Shengbing.
\newblock Effects of quarantine in six endemic models for infectious diseases.
\newblock {\em Mathematical Biosciences}, 2002.

\bibitem{Erdem:2017}
Mustafa Erdem, Muntaser Safan, and Carlos Castillo-Chavez.
\newblock {Mathematical Analysis of an SIQR Influenza Model with Imperfect
  Quarantine}.
\newblock {\em Bull Math Biol (2017) 79:1612–1636}, 79:1612–1636, 2017.

\bibitem{ElFatini:2020}
Mohamed~El Fatini, Roger Pettersson, Idriss Sekkak, and Regragui Taki.
\newblock {A stochastic analysis for a triple delayed SIQR epidemic model with
  vaccination and elimination strategies}.
\newblock {\em Journal of Applied Mathematics and Computing}, 64:781--805,
  2020.

\bibitem{Tagliazucchi:2020}
Tagliazucchi E., Balenzuela P., Travizano M., Mindlin~G. B., and Mininni~P. D.
\newblock {Lessons from being challenged by COVID-19}.
\newblock {\em Chaos, solitons, and fractals}, 137:109923, 2020.

\bibitem{Barrio:2009}
Barrio R.A, Varea C., Govezensky T., and Jos\'e M.V.
\newblock {Modelling the geographical spread of the influenza pandemic A(H1N1):
  The Case of Mexico.}
\newblock {\em Applied Mathematical Sciences}, 7:2143--2176, 2009.

\bibitem{Barrio:2020}
Barrio R.A., Kaski K.K., Haraldsson G.G., Aspelund T., and Govezensky T.
\newblock {Modelling COVID-19 epidemic in Mexico, Finland and Iceland}.
\newblock {\em arXiv}, page 2007.10806v1, 2020.

\bibitem{Bittihn:2020}
Bittihn P. and Golestanian R.
\newblock {Stochastic effects on the dynamics of an epidemic due to population
  subdivision}.
\newblock {\em Chaos}, 30:101102, 2020.

\bibitem{CDC:AUG2020}
Centers for Disease~Control and Prevention.
\newblock {\em {Duration of Isolation $\&$ Precaution for Adults with COVID-19,
  CDC, August 16th,2020}}, 2020.
\newblock \url{
  https://www.cdc.gov/coronavirus/2019-ncov/hcp/duration-isolation.html}.

\bibitem{Figar:2020}
S.~Figar, V.~Pagotto, L.~Luna, J.~Salto, M.~Wagner Manslau, A.~Mistchenko,
  A.~Gamarnik A. M.~Gomez Saldano, and F.~Quiros.
\newblock {Community-level SARS-CoV-2 Seroprevalence Survey in urban slum
  dwellers of Buenos Aires City, Argentina: a participatory research}.
\newblock {\em medRxiv}, 2020.

\bibitem{Eskild:2020}
Eskild Petersen, Marion Koopmans, Unyeong Go, Davidson~H Hamer, Nicola
  Petrosillo, Francesco Castelli, Merete Storgaard, Sulien~Al Khalili, and Lone
  Simonsen.
\newblock {Comparing SARS-CoV-2 with SARS-CoV and influenza pandemics}.
\newblock {\em Lancet. Infect. Dis.}, 20:E238--E244, 2020.

\bibitem{Oran:2020}
Daniel~P. Oran, AM, and Eric~J. Topol.
\newblock {Prevalence of Asymptomatic SARS-CoV-2 Infection}.
\newblock {\em Annals of Internal Medicine}, 173:362--367, 2020.

\bibitem{Alvin:2020}
J.~Alvin, Christine Cocks, and Jeffery~Peter Green.
\newblock {COVID-19: in the footsteps of Ernest Shackleton Thorax}.
\newblock {\em Annals of Internal Medicine}, 75:693--694, 2020.

\bibitem{Park:2020}
Park Y.J., Park O., Parkand S.Y., Kim Y.-M., Kim J., Kweon S., Woo Y.and~Gwack
  J., Kim S., Lee J., Hyun J., Ryu B., Jang Y.S., Kim H., Shin S.H., Yi~S., Lee
  S., Kim H.K., Lee H., Jin Y., Park E., Choi S.W., Kim M., Song J.and~Choi
  S.W., Kim D., Jeon B.H., Yoo H., and Jeong E.K.
\newblock {Contact tracing during coronavirus disease outbreak, South Korea}.
\newblock {\em Emerging Infectious Diseases}, 26:2465--2468, 2020.

\end{thebibliography}
\bibliographystyle{unsrt} 

\end{document}